\documentclass[aps,twocolumn,groupedaddress,floatfix,superscriptaddress,preprintnumbers]{revtex4}
\usepackage{amsmath,amssymb,multirow,epsfig,bm,pifont}
\usepackage[linkcolor=black,citecolor=black,urlcolor=blue,colorlinks=true,linktocpage=true]{hyperref}
\usepackage{tabularx}

\usepackage{ulem}

\newcommand{\beq}{\begin{equation}}
\newcommand{\eeq}{\end{equation}}
\newcommand{\bea}{\begin{eqnarray}}
\newcommand{\eea}{\end{eqnarray}}

\begin{document}

\preprint{
LA-UR-11-10456,
INT-PUB-11-015
}

\title{Exact-exchange density functional theory for neutron drops}
\author{Joaqu\'{\i}n E. Drut}
\affiliation{Theoretical Division, Los Alamos National Laboratory, Los Alamos, NM, 87545--0001, USA}
\affiliation{Department of Physics, The Ohio State University, Columbus, OH 43210--1117, USA}
\author{Lucas Platter}
\affiliation{Fundamental Physics, Chalmers University of Technology, SE-41296, G\"oteborg, Sweden}
\affiliation{Institute for Nuclear Theory, University of Washington, Seattle WA\ 98195 USA}
\affiliation{Department of Physics, The Ohio State University, Columbus, OH 43210--1117, USA}

\begin{abstract}
  We compute the ground-state properties of finite systems of
  neutrons in an external harmonic trap, interacting via the Minnesota potential,
  using the ``exact-exchange" form of orbital-dependent density
  functional theory.  We compare our results with Hartree-Fock 
  calculations and find very close agreement. Within 
  the context of the interaction studied, we conclude that this simple 
  orbital-dependent functional brings conventional nuclear density 
  functional theory to the level of Hartree-Fock in an {\it ab initio} fashion. 
  Our work is a first step towards higher-order {\it ab initio} nuclear 
  functionals based on realistic nucleon-nucleon interactions.
\end{abstract}

\date{\today}

\maketitle

\section{Introduction}
\label{sec:introduction}
Density functional theory (DFT) is a general theory of quantum
many-body systems with a long history. In its modern version, that 
began with the work of Hohenberg, Kohn and Sham~\cite{HohenbergKohn,
KohnSham}, DFT has become an essential tool in quantum chemistry as 
well as in materials science and condensed matter physics (see 
e.g.~\cite{DreizlerGross,ParrYang,Kohanoff}). 
This is largely due to the fact that large systems, intractable by {\it ab initio} methods
like Coupled Cluster (CC) or Quantum Monte Carlo (QMC) even with
modern computational power, are typically within the reach of
Kohn-Sham (KS) DFT with relatively modest computational
resources~\cite{KohnNobelTalk}. In spite of significant progress and
multiple successes, the fundamental challenge remains the same in all
applications: the central object of the theory, namely the energy
density functional (EDF) is a priori unknown and must be built either
phenomenologically or from first principles (or some combination
thereof) by implementing some kind of approximation scheme. 
Conventional DFT has slowly evolved from the first simple and
often uncontrolled approximation strategies of the early days, such as
Thomas-Fermi theory~\cite{ThomasFermi} and the Local Density Approximation 
(LDA)~\cite{KohnSham}, to more sophisticated semi-local approaches including the 
Generalized Gradient Approximation (GGA), meta-GGAs and hyper-GGAs~\cite{DFTLadder}.

Progress in the field of quantum chemistry~\cite{Engel} has pushed the boundaries of
such conventional approaches to DFT by allowing electronic EDFs to
depend explicitly on the single-particle orbitals of KS DFT. While the
set of possible functionals is thus enlarged, this additional freedom
comes at the price of increased formal complexity and computational
demand.  Indeed, the determination of the KS auxiliary one-body
potential $v^{}_\text{KS}$ in the case of orbital-dependent
functionals necessitates the solution of the Optimized Effective
Potential (OEP) integral equation~\cite{OEPEq,KurthPittalis}, which prevented 
widespread use of these functionals for a long time. Significant
advances in the last decade, however, have enabled the numerical
solution of the OEP equation in a more straightforward and systematic
fashion, thereby allowing orbital-dependent functionals to enter the 
mainstream of electronic DFT~\cite{Goerling,KP}.

These developments were often motivated by the inadequacies of
conventional electronic EDFs concerning some practical issues (such as
their inability to deal with Van der Waals forces, to predict the
existence of negative ions and to properly account for electron
self-interaction~\cite{Engel}) as well as formal problems (such as the lack of
particle-number derivative discontinuities~\cite{PerdewParrLevyBalduz}, 
inability to reproduce the correct long-range tail of the one-body potentials,
etc.~\cite{Engel}). Orbital-dependent functionals have not only
mitigated these issues, all of which stem from a poor description of particle 
exchange in GGA-based EDFs, but they have also cleared a path towards {\it
ab initio} DFT. Indeed, many-body perturbation theory expressions for the total
energy (at first order in its simplest form, i.e. Hartree-Fock, but
also at second and higher orders, and in resummed forms) are generally 
orbital-dependent functionals, which have a clear and direct connection to 
the microscopic Hamiltonian~\cite{BestOfBothWorlds}.

Many-body systems of electrons in external fields are the standard
arena for the application of DFT. However, nuclear physics shares the
same interests in many-body techniques and faces very similar
challenges. DFT has therefore become a standard tool for the
computation of the properties of heavy nuclei. In nuclear DFT, the electrons 
are replaced with nucleons, and the nucleon-nucleon interaction plays the role of the 
Coulomb interaction. The vast majority of
nuclear functionals, whether phenomenologically motivated or derived
from first principles, are largely in the category of GGAs (broadly defined).
The parameters in these functionals are usually fit to a subset of stable
nuclei (with the exception of some of the latest functionals, such as SLy4 or SkO,
which have been fit to some experimentally accessible unstable nuclei)
and are therefore not directly related to our understanding of the 
nucleon-nucleon interaction.

The main objective of this work is to explore the prospects of ab
initio DFT for the nuclear case. A more complete survey of the current status
of the field can be found in Ref.~\cite{DFP}. In this work, we implement one of the 
simplest possible orbital-dependent DFTs derived from an underlying hamiltonian, 
the so-called ``exact-exchange" (EXX) form. We apply this to the case of neutron 
drops interacting via the Minnesota potential~\cite{MinnPot}. We then compare the 
results of this DFT with exact Hartree-Fock (HF) for various numbers of neutrons
in a harmonic external potential.

The EXX functional is easily defined: it consists of the HF energy, in
which the single-particle HF orbitals are replaced with the KS
orbitals. Thus, the fundamental difference between EXX DFT and exact
HF is that at each step in the iterative optimization procedure the HF
approach involves a {\it non-local} auxiliary potential, whereas KS
DFT (regardless of the form of the functional) utilizes a {\it local}
auxiliary potential $v^{}_\text{KS}$. (It remains an open question whether
such local auxiliary field will be too constraining for nuclear physics, see e.g. Ref.~\cite{DuguetLesinski}.)
In this sense, EXX DFT can be
regarded as a constrained optimization of the HF energy, where the
constraint consists in demanding the locality of $v^{}_\text{KS}$. As
a consequence, the ground-state energies obtained via EXX DFT should be
expected to be higher than those of HF.  As we shall see, the
differences are very small in the case we study, but this will in
general depend on the form of the interaction.

While EXX DFT is formally simple, we do not claim that it is accurate
in an absolute sense, but rather that it represents an extremely
accurate approximation to exact HF. As mentioned above, this property
will in general depend on the interaction. However, we wish to stress
this point as a promising feature of orbital-dependent DFT, even
though HF calculations are well known to be a poor description of
nuclei due to the non-perturbative nature of the nuclear interaction
at short distances (c.f. high momenta). A more realistic description
must both transcend perturbation theory and include pairing
correlations at least at the mean-field level, i.e. \`a la
Hartree-Fock-Bogoliubov (HFB). Work in this direction has recently
involved efforts towards taming the nuclear interaction at high
momenta using Renormalization Group (RG) transformations~\cite{SRG_1}
\footnote{A possible alternative approach would be to employ directly chiral
  effective theory interactions that are soft by construction
  \cite{Coraggio:2007mc}}. These are transformations that leave
observables unchanged but reduce the strength of the potential at
large momenta. Such transformations render the problem more
perturbative while maintaining the hierarchy of many-body forces. In
this context, our work may be regarded as a necessary first step in an
{\it ab initio} DFT program that connects microscopic RG-transformed
Hamiltonians with orbital-dependent DFT based on HFB plus perturbation
theory.


\section{Kohn-Sham DFT and the Optimized Effective Potential}
\label{sec:kohn-sham-dft}
In quantum chemistry, the ``optimized potential
method", or simply the ``optimized effective potential", refers
collectively to the use of orbital-dependent EDFs and to the
determination of the KS auxiliary potential $v^{}_\text{KS}$ by
solving the OEP integral equation. For completeness, and in order to
set our notation, we present here a short derivation of this equation,
along with a brief review of KS DFT. For simplicity, we shall
restrict ourselves to functionals that do not depend on the KS
eigenvalues; the corresponding generalization is easy to carry out.

\subsection{Derivation of the OEP equation}
\label{sec:deriv-oep-equat}
The central tenet of DFT is the Hohenberg-Kohn (HK) theorem, whereby
the existence of an energy density functional is asserted, of the form
\beq
E[\rho] = F[\rho] + E_\text{ext}[\rho],
\eeq
where
\beq
E_\text{ext}[\rho] = \int d{\bf x} \ v^{}_\text{ext}({\bf x}) \rho({\bf x}),
\eeq
such that $F[\rho]$ depends only on the one-body density $\rho({\bf x})$. 
One may think of the latter as the total density, but in general it may denote 
spin, isospin, kinetic or anomalous densities, in which case $\bf x$ represents 
a collective index for the coordinate and every other degree of freedom.

The external potential $v^{}_\text{ext}$ represents the electric
field of the ions, and it confines the system to a particular spatial
region. This is a fundamental difference between the electronic and
nuclear cases, since in the latter the system is self-bound,
i.e. there is no $v^{}_\text{ext}$~\cite{DFP}.

According to the HK theorem, the functional form of $F$ is determined
solely by the interactions and not by the external potential
$v^{}_\text{ext}$; the functional $F$ is therefore said to be 
{\it universal}. The HK theorem is an existence theorem and gives therefore
no instructions as how to build or find this functional.
Kohn-Sham DFT takes a first formal step towards the explicit
construction of $F[\rho]$ by separating it into a non-interacting
piece, i.e. the kinetic energy of the free system, and everything
else:
\beq
\label{eq:functional}
F[\rho] = T_s + E^{}_\text{int}[\rho],
\eeq
where
\beq
\label{eq:kinetic}
T_s = \sum_{\sigma=\uparrow,\downarrow}\sum_{k=1}^{N_\sigma} \int d{\bf x} \ \varphi^{*}_{k\sigma}({\bf x}) \left ( -\frac{\hbar^2\nabla^2}{2m} \right ) \varphi^{}_{k\sigma}({\bf x}) .
\eeq
It should be stressed that $T_s$ represents the kinetic energy of the
auxiliary KS system, which in general is different from that of the
many-body system. Their difference is assumed to be accounted for in
$E^{}_\text{int}[\rho]$.

The $\varphi^{}_{k\sigma}({\bf x})$ are a set of auxiliary single-particle orbitals (the KS orbitals), such that
$\rho = \rho^{}_\uparrow + \rho^{}_\downarrow$, where
\beq
\label{rhodef}
\rho^{}_\sigma({\bf x}) = \sum_{k=1}^{N_\sigma} |\varphi_{k\sigma}({\bf x})|^2~,
\eeq
and $N^{}_\sigma$ is the particle number for spin $\sigma$.  Kohn-Sham DFT then proceeds
to optimize the energy functional by solving a Schr\"odinger-like
equation:
\beq
\label{KSeq}
\left [ -\frac{\hbar^2\nabla^2}{2m}  + v^{}_{\text{KS},\sigma}({\bf x}) \right ]\varphi^{}_{k\sigma}({\bf x}) = \epsilon^{}_k \varphi^{}_{k\sigma}({\bf x}),
\eeq
for all $k$. As mentioned above, this should be contrasted with the HF
approximation, in which the corresponding Schr\"odinger equation
involves a non-local potential. In this sense, EXX DFT can be regarded
as resulting from an HF minimization procedure, with the added
constraint that the auxiliary potential be local. In spite of this
constraint, orbital-dependent DFT beyond EXX has the potential to
surpass HF, and in fact does so in practice in the electronic case (see e.g. Ref.~\cite{Engel}).

By definition, the KS potential in Eq.~(\ref{KSeq}) is given by
\beq
\label{vKSdef}
v^{}_{\text{KS},\sigma}({\bf x}) \equiv \frac{\delta V}{\delta \rho^{}_\sigma({\bf x})} 
= v^{}_\text{ext}({\bf x}) + \frac{\delta E^{}_\text{int}}{\delta \rho^{}_\sigma({\bf x})},
\eeq
where
\beq
V \equiv E^{}_\text{ext} + E^{}_\text{int}.
\eeq
The second term in the r.h.s. of Eq.~(\ref{vKSdef}) will in general
depend on the KS orbitals, such that Eq.~(\ref{KSeq}) is to be solved
self-consistently by starting with a guess for the orbitals or for
$v^{}_\text{KS}$.

It is at this point that orbital-dependent DFT departs from GGA DFT,
in which functionals depend {\it explicitly} on the density and its
gradients. Indeed, once we allow $E^{}_\text{int}$ to depend {\it
  explicitly} on the KS orbitals, it becomes unclear how to determine
the KS potential using its definition Eq.~(\ref{vKSdef}).  One of the
simplest ways to proceed is to use the chain rule of functional
differentiation and consider the following identity
\beq
\label{dExcdv_A1}
\frac{\delta V} {\delta v^{}_{\text{KS},\sigma}({\bf x})} = 
\sum_{\sigma'=\uparrow,\downarrow}\sum_{k=1}^{N_{\sigma'}} \int d{\bf y} \ \frac{\delta V}{\delta \varphi^{}_{k\sigma'}({\bf y})}
\frac{\delta \varphi^{}_{k\sigma'}({\bf y})}{\delta v^{}_{\text{KS},\sigma}({\bf x})} + \rm{c.c.}
\eeq 
In order to proceed we vary both sides of
the eigenvalue equation Eq.~(\ref{KSeq}) (and its complex conjugate):
\bea
\frac{\delta \varphi^{}_{k\sigma'}({\bf y})}{\delta v^{}_{\text{KS},\sigma}({\bf x})} =  G^{\sigma'\sigma}_{k}({\bf y},{\bf x}) \varphi^{}_{k\sigma}({\bf x}),
\eea
where
\beq
G^{\sigma' \sigma}_k({\bf y},{\bf x}) = \delta_{\sigma\sigma'} G^{}_{k\sigma}({\bf y},{\bf x}),
\eeq
with
\beq
G^{}_{k\sigma}({\bf y},{\bf x}) = 
\sum_{q\neq k} \frac{\varphi^*_{q\sigma}({\bf x})\varphi^{}_{q\sigma}({\bf y})}{\epsilon^{}_k - \epsilon^{}_q},
\eeq
is the Green's function. Equation~(\ref{dExcdv_A1}) then becomes
\beq
\label{dExcdv_A2}
\frac{\delta V} {\delta v^{}_{\text{KS},\sigma}({\bf x})} = 
\sum_{k=1}^{N_\sigma} \int d{\bf y} \ 
\frac{\delta V}{\delta \varphi^{}_{k\sigma}({\bf y})} G^{}_{k\sigma}({\bf y},{\bf x}) \varphi^{}_{k\sigma}({\bf x}) + \rm{c.c.}
\eeq

On the other hand, the HK theorem allows us to take implicit 
derivatives with respect to the density $\rho^{}_{\sigma}({\bf x})$:
\bea
\label{dExcdv_B}
\frac{\delta V}{\delta v^{}_{\text{KS},\sigma}({\bf x})} \!&=&\!
\int d {\bf x}_1 \frac{\delta V}{\delta \rho^{}_\sigma({\bf x}_1)} \frac{\delta \rho_\sigma({\bf x}_1)}{\delta v^{}_{\text{KS},\sigma}({\bf x})} \\
&&\!\!\!\!\!\!\!\!\!\!\!\!\!\!\!\!\!\!\!\!\!\!\! = 
\int d {\bf x}_1 v^{}_{\text{KS},\sigma}({\bf x}_1) \frac{\delta \rho_\sigma({\bf x}_1)}{\delta v^{}_{\text{KS},\sigma}({\bf x})} \nonumber \\
&&\!\!\!\!\!\!\!\!\!\!\!\!\!\!\!\!\!\!\!\!\!\!\! =
\sum^{N_\sigma}_{k=1} \int d {\bf x}_1 v^{}_{\text{KS},\sigma}({\bf x}_1) {\varphi^{*}_{k\sigma}({\bf x}_1)} G^{}_{k\sigma}({\bf x}_1,{\bf x}) \varphi^{}_{k\sigma}({\bf x}) \nonumber \\
&& + \ \ \rm{c.c.},\nonumber
\eea
where we have used Eqs.~(\ref{rhodef}) and~(\ref{vKSdef}).
Combining Eqs.~(\ref{dExcdv_A2}) and (\ref{dExcdv_B}) we arrive at 
\beq
\label{OEP_Eq}
\sum^{N_\sigma}_{k=1}\left( \psi^*_{k\sigma}({\bf x}) \varphi^{}_{k\sigma}({\bf x})+\rm{c.c.}\right) = 0,
\eeq
where we have defined the ``orbital shift" $\psi^{}_{k\sigma}$ by
\bea
\label{OrbitalShift}
\psi^*_{k \sigma} ({\bf x}) &\equiv& \int\! d {\bf x}_1 \left [ \frac{\delta V}{\delta \varphi^{}_{k \sigma}({\bf x}_1)}
- v^{}_{\text{KS},\sigma}({\bf x}_1) \varphi^{*}_{k\sigma}({\bf x}_1) \right ] \nonumber \\
&&\times \ G^{}_{k \sigma}({\bf x}_1,{\bf x}).
\eea
%
%
%
%
Equation (\ref{OEP_Eq}) is the OEP integral equation that defines the KS potential of orbital-based DFT.

\subsection{Solving the OEP equation}

While there are in principle multiple ways to solve the OEP equation,
we have found the one originally due to K\"ummel and Perdew~\cite{KP}
to be particularly useful. We shall refer to this method as the
K\"ummel-Perdew (KP) algorithm. This algorithm is iterative and starts
by noting that the orbital shifts can be found by solving a
differential equation, namely
\beq
\label{OrbShift_Eq}
\left [ -\frac{\hbar^2\nabla^2}{2m}  + v^{}_{\text{KS},\sigma}({\bf x}) - \epsilon^{}_k \right ] \psi_{k\sigma}^*({\bf x}) = 
\Lambda^{}_{k\sigma}({\bf x}) \varphi^{*}_{k\sigma}({\bf x})
\eeq
and
\bea
\!\!\!\!\Lambda^{}_{k\sigma}({\bf x}) \!&\equiv&\! v^{}_{\text{KS},\sigma}({\bf x}) - u^{}_{k\sigma}({\bf x}) \nonumber \\
&&\!\!\!\!\!\!\!\!\!\!\!\!\!\!\!-\!\! \int d {\bf x}^{}_1 (v^{}_{\text{KS},\sigma}({\bf x}^{}_1 )\!-\! u^{}_{k\sigma}({\bf x}^{}_1 )) 
\varphi^{*}_{k\sigma}({\bf x}^{}_1) \varphi^{}_{k\sigma}({\bf x}^{}_1 ),
\eea
with 
\beq
\label{ukDef}
u^{}_{k\sigma}({\bf x}) \equiv \frac{1}{\varphi^*_{k\sigma}({\bf x})}\frac{\delta V}{\delta \varphi^{}_{k\sigma}({\bf x})},
\eeq
where we assume some form for the initial $v^{}_{\text{KS},\sigma}({\bf x})$. Once the orbital shifts are determined, 
$v^{}_{\text{KS},\sigma}$ is updated according to
\beq
v^\text{new}_{\text{KS},\sigma}({\bf x}) = v^\text{old}_{\text{KS},\sigma}({\bf x}) + cS^{}_\sigma({\bf x}),
\eeq
where $c$ is a positive constant and the function
\beq
S^{}_\sigma({\bf x}) \equiv \sum^{N_\sigma}_{k=1}
\left( \psi_{k\sigma}^*({\bf x}) \varphi^{}_{k\sigma}({\bf x})+\rm{c.c.}\right)
\eeq
(c.f. Eq.(\ref{OEP_Eq})) is used as a local measure of the deviation
of the current $v^{}_{\text{KS}}$ with respect to the true solution of the
OEP equation. Indeed, wherever $v^{}_{\text{KS},\sigma}({\bf x})$ differs from the
true OEP, $S^{}_\sigma({\bf x})$ will be non-zero. The updated $v^{}_{\text{KS}}$ can
be used to re-compute the orbital shifts, keeping the KS orbitals and
eigenvalues constant, until the desired convergence criterion is
satisfied. The final $v^{}_{\text{KS}}$ is then re-inserted in the KS
equation (\ref{KSeq}) to compute a new set of orbitals, closing the
self-consistency loop.

It is easy to see that Eq.~(\ref{OrbShift_Eq}) is singular. Adding
any multiple of $\varphi^{}_{k\sigma}({\bf x})$ to $\psi_{k\sigma}^*({\bf x})$ gives a
new non-trivial solution. As first noted in Ref.~\cite{KP}, however,
Eq.~(\ref{OrbShift_Eq}) can still be solved in the subspace of
interest by using the method of Conjugate Gradients~\cite{CG}. Further
details of this approach can be found in Ref.~\cite{KP}.

The determination of the orbital shifts $\psi_{k\sigma}^*$ via
Eq.~(\ref{OrbShift_Eq}) represents an additional computational cost
that is not present in LDA- or GGA-based DFT. Indeed, in the latter
$v^{}_{\text{KS},\sigma}$ is determined simply by inserting the density and its
gradients in a pre-calculated analytic expression derived from the EDF
via Eq.~(\ref{vKSdef}). In contrast, in the OEP method (within the KP
approach) the calculation of all the $\psi_{k\sigma}^*$'s necessitates the
solution of $N$ differential equations, where $N$ is the number of
particles. While this is a considerable increase in computational
demand, it should be kept in mind that the $\psi_{k\sigma}^*$'s are
independent from each other, such that the $N$ differential equations
can be solved in a parallel fashion with perfect scaling (up to
communication costs required to broadcast $v^{}_{\text{KS}}({\bf x})$ at the
beginning and to compute $S^{}_\sigma({\bf x})$ at the end).

\section{EXX energy density functional and the Minnesota potential}
\label{sec:minnesota-potential}
We have used in this work the Minnesota nucleon-nucleon interaction of
Ref.~\cite{MinnPot}, which is given for pure neutron systems by
\bea
V(r) &=&
\left( V_{\text R}(r)
+  V^{}_{\text s}(r) \mathcal{P}_s
+ V^{}_{\text t}(r) \mathcal{P}_t\right)
\nonumber \\
&& \times
\frac{1}{2}\left ( 1 +  P^r \right )~,
\eea
%
where $\mathcal{P}_s$ ($\mathcal{P}_t$) is the operator that projects
on the spin singlet (triplet) state and $P^r$ is the coordinate
exchange operator. The potentials in the various channels have
Gaussian forms given by
\bea
\label{eq:minnesota}
V_{\text R}  &=& V_{\text{0R}}  \exp(-\kappa_{\text{R}} r^2_{}) \nonumber \\
V_{\text t}  &=& -V_{\text{0t}} \exp(-\kappa_{\text{t}} r^2_{})  \nonumber \\
V_{\text s}  &=& -V_{\text{0s}} \exp(-\kappa_{\text{s}} r^2_{}),
\eea
 and
\bea
V_{\text{0R}} &=& 200.0\ \text{MeV}, \ \ \ \ \  \kappa_{\text{R}} = 1.487\ \text{fm}^{-2} \nonumber  \\
V_{\text{0t}} &=& 178.0\ \text{MeV}, \ \ \ \ \  \kappa_{\text{t}} = 0.639\ \text{fm}^{-2} \nonumber  \\
V_{\text{0s}}&=& 91.85\ \text{MeV}, \ \ \ \ \  \kappa_{\text{s}} = 0.465\ \text{fm}^{-2}.
\eea
Our main reason for using this potential is that it provides 
a semi-realistic yet easy-to-implement interaction. In addition, this 
potential is moderately soft and as such it is comparable to those 
obtained using renormalization-group methods~\cite{SRG_1}. Other 
approaches to {\it ab initio} DFT are also being explored using this simple 
potential, and No-Core-Full-Configuration (NCFC) and Coupled-Cluster (CC) 
results have recently become available as well, which makes this a good test 
case for OEP methods.

The energy density functional is of the form given in
Eq.~\eqref{eq:functional} and the mass in the kinetic energy term
(Eq.~\eqref{eq:functional}) has been set to $m_n = 939$ MeV. In
practice, one combines the kinetic term with the external Harmonic Oscillator
(HO) potential energy. This is convenient because the KS orbitals can
be expanded using the HO basis $\{\phi_j^{}\}$,
\bea
\varphi_{k\sigma}({\bf x}) &=& \sum _{j=1}^{N_\text{max}} a^{}_{jk,\sigma} \phi_j^{}({\bf x}) \\
a^{}_{jk,\sigma} &=& \int d{\bf x}\ \phi_j^{*}({\bf x}) \varphi^{}_{k\sigma}({\bf x}),
\eea
and in this basis the sum $T_s + E_\text{ext}$ takes a simple diagonal form.

In the EXX case considered in this work the interaction enters through
\beq
E_\text{int} = 
\frac{1}{2} \sum_{ijkl}{\bar V}_{ijkl}\rho^{}_{ki}\rho^{}_{lj},
\eeq
where we have used collective indices $i$ to denote the pair
$(i,\sigma_i)$ for the basis element and the spin. Here,
\beq
\rho^{}_{ij} \equiv \rho^{}_{i\sigma_i;j\sigma_j} = \delta_{\sigma_i \sigma_j} \sum_{k=1}^{N_\sigma} a^{*}_{ik,\sigma_i} a^{}_{jk,\sigma_i}
\eeq
is the one-body density matrix, and
\beq
{\bar V}^{}_{ijkl} = {V}^{}_{ijkl} - {V}^{}_{ijlk}
\eeq
are the anti-symmetrized matrix elements of the interaction. As
mentioned above, this form of $F[\rho]$ is simply that of the HF
energy, but with the KS orbitals replacing the HF single-particle
wavefunctions. The interaction in Eq.~\eqref{eq:minnesota} is local
and we therefore obtain
\bea
{V}^{}_{ijkl} = 
\int d{\bf x}^{}_1 d{\bf x}^{}_2 \phi^{*}_{i}({\bf x}^{}_1)
\phi^{*}_{j}({\bf x}^{}_2)
V(r)
\phi^{}_{k}({\bf x}^{}_1)
\phi^{}_{l}({\bf x}^{}_2),
\eea
%
where $r = |{\bf x}^{}_1 - {\bf x}^{}_2|$ and the total spin of the
initial ($ik$) and final ($jl$) states is the same.

%
%
%

\section{Results and Conclusions}

We have considered systems of $8$ and $20$ neutrons for various trap
frequencies and basis sizes, and computed the total energy
\beq
E_ \text{tot}^{} = E_\text{kin}^{} + U_\text{ext} + E_\text{H}^{} + E_\text{F}^{},
\eeq
where $E_\text{kin}^{} = T_s^{}$, and
\beq
U_\text{ext} = \int d {\bf x}\ v_\text{ext}^{}({\bf x}) \rho({\bf x}),
\eeq
where
\beq
v_\text{ext}^{}({\bf x}) = \frac{1}{2} m_n^{} \Omega^2{\bf x}^2,
\eeq
is the HO external potential.

The total energy can also be obtained from eigenvalue sum rules, which
we have verified numerically, and are given in the HF case by
\beq
E_\text{tot}^{} = \sum^{}_{k,\sigma} \epsilon_k - E^{}_H - E^{}_F,
\eeq
where each term is evaluated using HF orbitals, and in the EXX-DFT case by
\beq
E_\text{tot}^{} = \sum^{}_{k,\sigma} \epsilon_k + E^{}_H + E^{}_F - 
\sum_{\sigma} \int d{\bf x} \ v^{}_{\text{KS},\sigma}({\bf x}) \rho_\sigma({\bf x}),
\eeq
where the EXX-DFT orbitals should be used. Notice that in both sum
rules the eigenvalue sums go over both spins.  The Hartree and Fock
energies are respectively given by
\beq
E_\text{H}^{} = \frac{1}{2} \sum_{ijkl} V^{}_{ijkl}\rho^{}_{ki}\rho^{}_{lj}
\eeq
\beq
E_\text{F}^{} = -\frac{1}{2} \sum_{ijkl} V^{}_{ijlk}\rho^{}_{ki}\rho^{}_{lj}.
\eeq

We have also computed the r.m.s. radius $\sqrt{\langle r^2 \rangle}$, where
\beq
\langle r^2 \rangle = \frac{1}{N}  \int d{\bf x} \ r^2 \rho(r)~,
\eeq
as well as the form factor, given by
\beq
F(q) =  4\pi \int \! dr\; r^2  \rho(r) \frac{\sin q r}{q r},
\eeq
where $q = |{\bf q}|$ and $r = |{\bf x}|$.

\begin{figure}[b] 
\epsfig{file=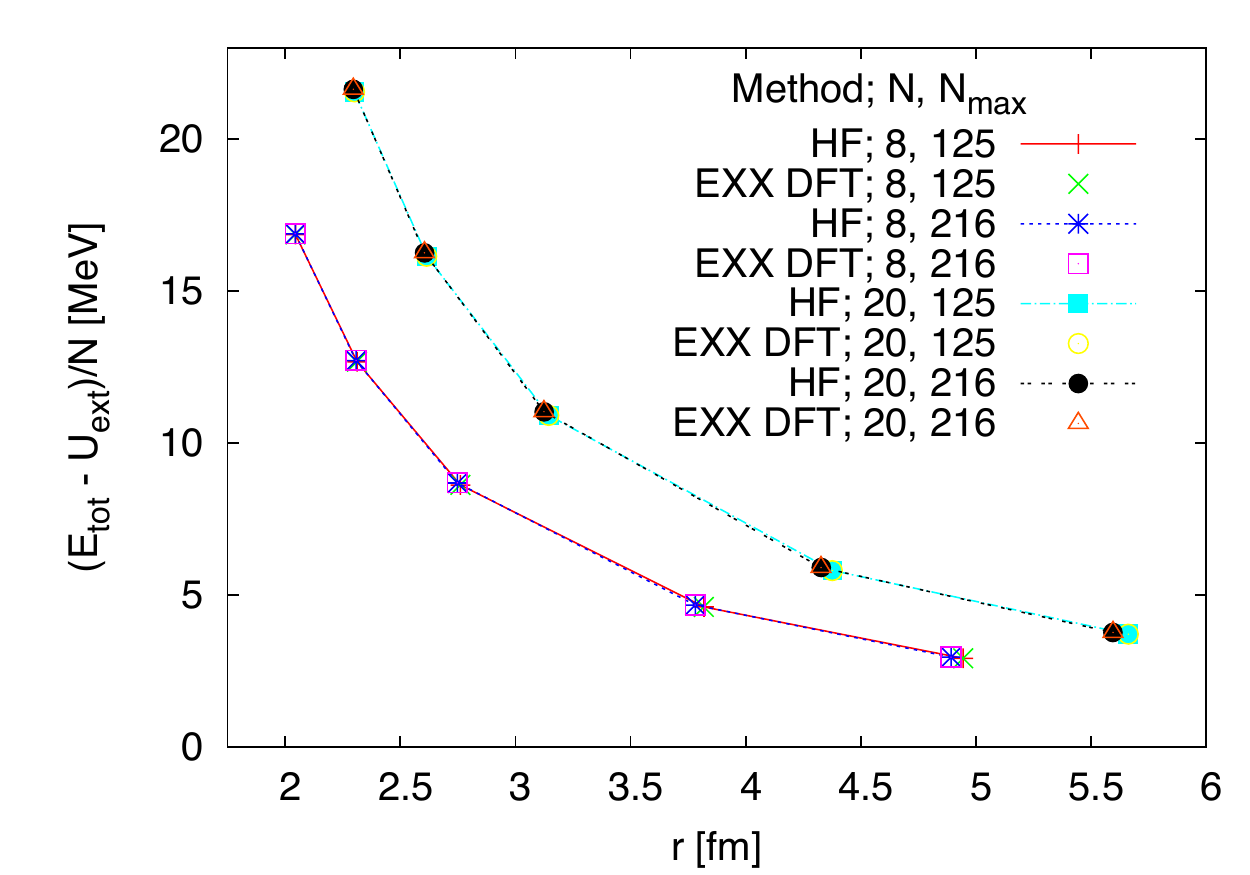, width=\columnwidth}
\caption{(Color online) Internal energy as a function of the r.m.s. radius for 8 (lower curve) and 20 (upper curve) neutrons, 
computed with HF and EXX DFT, for two different basis sizes, namely $N_\text{max}^{}=$125 and 216. 
From left to right, the trapping potential corresponds to $\hbar \Omega \!=\!$ 20, 15, 10, 5 and 3 MeV.}
\label{Fig:EvsR}
\end{figure}

\begin{figure}[t] 
\epsfig{file=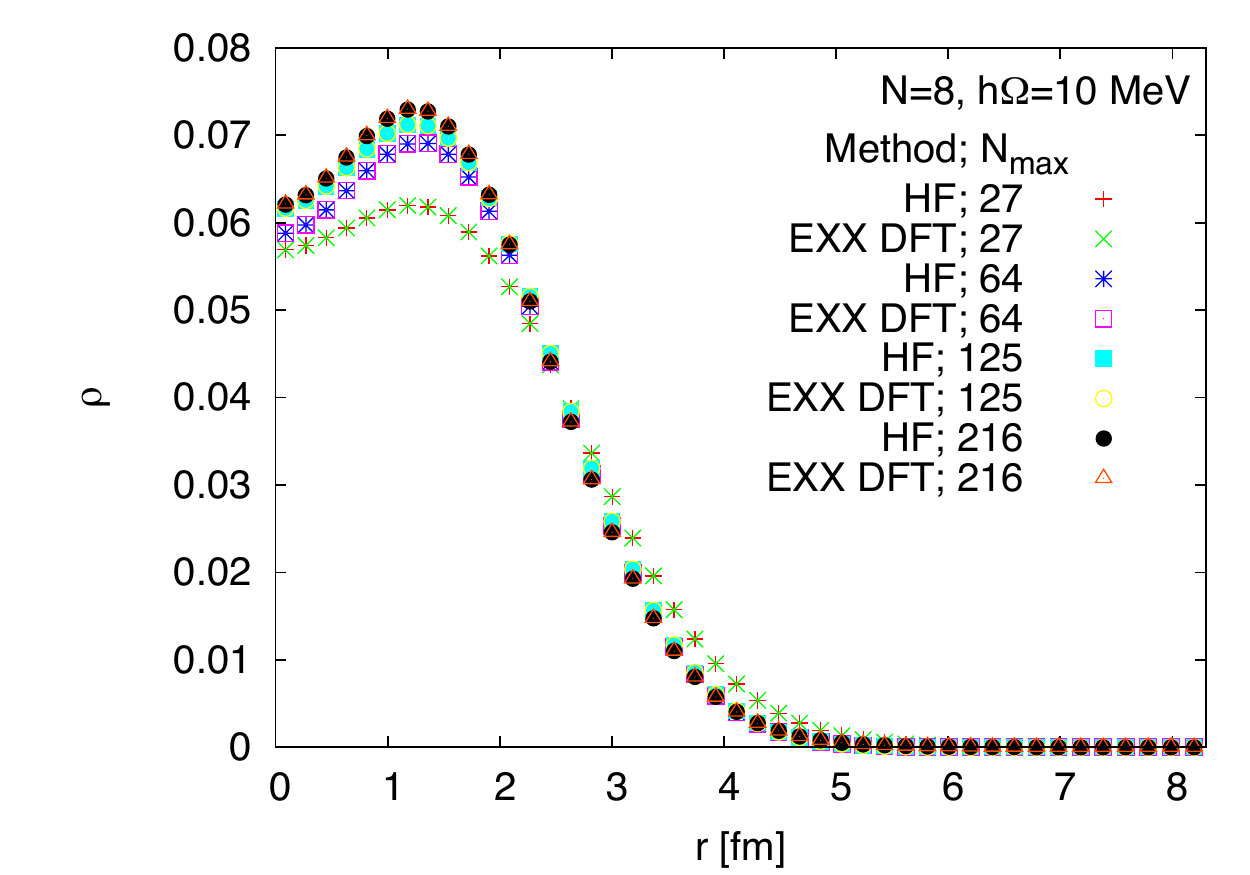, width=\columnwidth}
\caption{(Color online) Density profiles of a system of 8 neutrons in a $\hbar \Omega \!=\! 10$ MeV trap as a function of 
the basis size $N_\text{max}^{}$, computed with HF and EXX DFT.}
\label{Fig:DensityProfiles}
\end{figure}

\begin{figure}[t] 
\epsfig{file=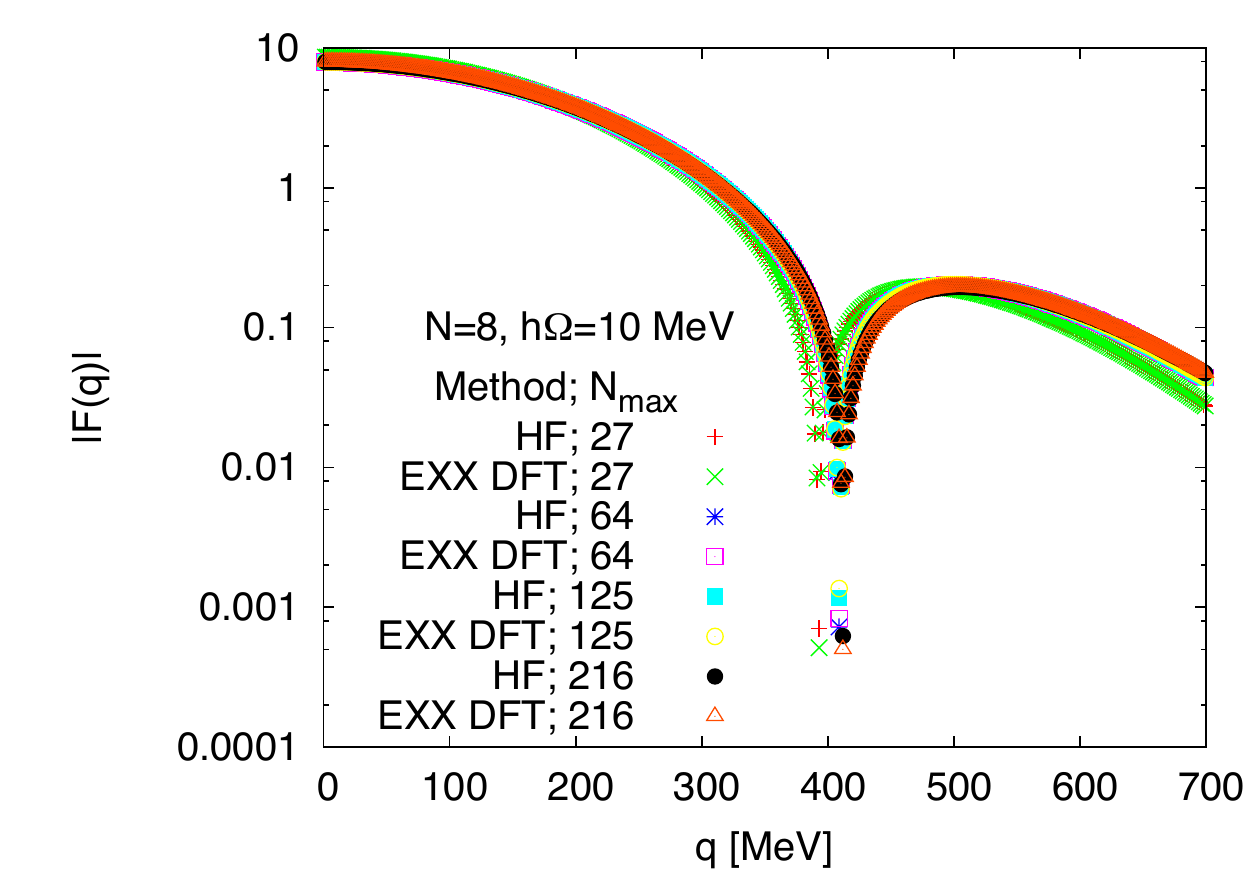, width=\columnwidth}
\caption{(Color online) Form factor of a system of 8 neutrons in a $\hbar \Omega \!=\! 10$ MeV trap as a function of 
the basis size $N_\text{max}^{}$, computed with HF and EXX DFT.}
\label{Fig:FormFactor}
\end{figure}

Our results for the internal energy $E_{tot}^{} - U_\text{ext}$ per
particle, as a function of the r.m.s. radius, are shown in
Fig.~\ref{Fig:EvsR}, for various frequencies of the trapping
potential. Also shown in Fig.~\ref{Fig:EvsR} are the results corresponding to
$8$ and $20$ neutrons, and in all cases we display the degree of
convergence of HF and EXX DFT, with respect to the size of the basis 
$N_\text{max}^{}$, by showing data for $N_\text{max}^{}=$ 64, 125 and 216.
Figures~\ref{Fig:DensityProfiles} and~\ref{Fig:FormFactor} show the
density profiles and the corresponding form factors for a fixed
external harmonic potential of frequency $\hbar \Omega \!=\! 10$ MeV and
basis size $N_\text{max}^{}=$ 27, 64, 125 and 216. Throughout this work we 
have chosen the value of $\hbar \Omega$ for the basis equal to that of the trapping potential.
As seen in the plots, in all cases the convergence pattern of HF and EXX DFT 
as a function of $N_\text{max}^{}$ is the same. At fixed $N_\text{max}^{}$,
on the other hand, the convergence patterns as a function of the
respective HF and KS iterations are significantly different from each
other. We provide a partial summary of our results in Table~\ref{TableEnergies}.

We have implemented a parallel code (using OpenMP) for both the HF and
OEP calculations. The latter use some of the HF routines and a set of
extra ones to solve the OEP equations.  Using 8 processors and some
limited optimizations of the parameters (such as the constant $c$ in
the KP algorithm and the number of iterations in the KP loop) we have
found that, at fixed accuracy for the convergence criterion, OEP
calculations require about a factor of $2-3$ more iterations than HF,
and each OEP iteration takes about a factor of $2-3$ more time than
its HF counterpart, for the basis sizes and particle numbers studied
here. Thus, overall the OEP method at the EXX-DFT level is a factor of
$4-9$ slower than HF.  However, further optimizations are possible if
for example MPI is used in the parallelization of the calculation of
the orbital shifts. This would have significant impact in calculations
for larger particle number and larger basis size, as well as for more
sophisticated functionals.


\begin{table}[b]
\begin{center}
\caption{
\label{TableEnergies}
Summary of results for the energies in MeV for Hartree-Fock (HF) and Exact-Exchange DFT (EXX),
broken up into total ($E^{}_ \text{tot}$), kinetic ($E^{}_\text{kin}$), Hartree ($E^{}_\text{H}$), 
Fock ($E^{}_\text{F}$), and internal ($E^{}_ \text{I} = E^{}_\text{tot} - U^{}_\text{ext}$). Also shown is the r.m.s. 
radius $\sqrt{\langle r^2\rangle}$.
}
\begin{tabularx}{\columnwidth}{@{\extracolsep{\fill}}c c c c c c c c c}
       \hline
       $Method$ & $N$ & $\hbar \Omega$ & $E^{}_ \text{tot}$& $E^{}_\text{kin}$ & $E^{}_\text{H}$ & $E^{}_\text{F}$ &
       $E^{}_ \text{I}$ & $\sqrt{\langle r^2\rangle}$ \\
       \hline \hline
       HF      &       8       &       20                      & 296.36 &  200.92 & -225.18 &  159.35 & 135.09 &  2.0438       \\
       EXX     &       8       &       20              & 296.36 &  200.99 & -225.24 &  159.39 & 135.14 &  2.0435       \\

       \hline
       HF      &       8       &       10                      &  142.43 &  111.22 & -125.38 &  83.70 & 69.53 & 2.7484 \\
       EXX     &       8       &       10              &  142.44 &  111.27 & -125.44 &  83.74 & 69.56 & 2.7478 \\

       \hline
       HF      &       8       &       3                       &  44.51 &  35.18 & -30.51 &  19.05 &  23.72  & 4.8927          \\
       EXX     &       8       &       3               &  44.51 &  35.19 & -30.52 &  19.05 &  23.72  & 4.8921          \\

       \hline
       HF      &       20      &       20                      &  941.75 &  707.69  & -920.03 &  645.08 &  432.74 & 2.2965     \\
       EXX     &       20      &       20              &  941.77 &  707.83  & -920.19 &  645.20 &  432.84 & 2.2963     \\

       \hline
       HF      &       20      &       10                      &  456.25  & 382.73  & -488.61 &  326.55 &  220.68  & 3.1246    \\
       EXX     &       20      &       10              &  456.26  & 382.88  & -488.80 &  326.69 &  220.77  & 3.1240 \\

       \hline
       HF      &       20      &       3                       &  143.52 &  119.55 &  -118.19 &   74.15 &  75.52 &  5.5958     \\
       EXX     &       20      &       3               &  143.52 &  119.57 &  -118.21 &   74.17 &  75.53 &  5.5954     \\

\hline
\end{tabularx}
\end{center}
\end{table}

\section{Summary and Outlook}
\label{sec:summary-outlook}
In this work we have computed the energy, radius, density profile and
form factor of finite systems of neutrons in a harmonic trap. For this
purpose we have used pure HF and orbital-based DFT in the EXX form,
implementing the algorithm of K\"ummel and Perdew to solve the OEP
equation. Our results for the OEP approach agree at the $0.1\%$ level
or better with those of pure HF for all the quantities we computed.
We note in particular that such an agreement was reached for each
basis size even though HF and OEP converge to their respective results
in completely different ways. Although limited in scope, our work
shows that it is possible to capture the exchange aspects of the
nuclear interaction in an extremely accurate fashion completely within
the context of local KS-DFT.

As mentioned in the introduction, this work represents the first step
in a program whose objective is to construct a nuclear energy density 
functional from first principles, employing orbital-based methods. 
Ideally, this functional would be able to at least qualitatively predict the properties of 
heavy nuclei and would only be based on our microscopic understanding of 
the internucleon interaction. We do not exclude the possibility, however, that 
an accurate quantitative description of energies and radii may require readjusting 
certain parameters in this functional. In this sense, our work is a step forward 
from LDA- and GGA-based DFT, which are typically phenomenological in nature 
in the nuclear case, and therefore limited in their ability to predict the properties of
unknown systems.  Much work remains to be done, however, to bring this
approach to a level of accuracy that is competitive with state-of-the-art nuclear EDFs.

Orbital-based DFT is one possible road out among several
currently pursued in the nuclear DFT community. An alternative way is to find a
controlled scheme that approximates the non-local ab initio functional
with a local one and thereby facilitates the application of standard
KS DFT (see e.g. \cite{DME}).

The next step in this {\it ab initio} DFT program will be to treat nuclei
by implementing realistic nuclear interactions such as
chiral interactions softened with modern renormalization group
methods. This will require extending the current framework to
include many-body forces and non-central as well as non-local terms in
the interaction. This is currently underway; we have successfully performed 
proof-of-principle calculations with three-body forces, and shown that the formalism
for many-body forces at the EXX level is a straightforward extension of our
presentation in Sec.~\ref{sec:kohn-sham-dft} \cite{DrutPlatter}. Extensions of the 
OEP method currently in progress involve improved functionals, e.g. those generated 
by second-order perturbation theory, as well as modifications to account for 
pairing correlations.


\begin{acknowledgments}

We acknowledge support under U.S. DOE Grants No.~DE-FG02-00ER41132 and DE-AC02-05CH11231, 
UNEDF SciDAC Collaboration Grant No.~DE-FC02-07ER41457 and NSF Grant No.~PHY--0653312
and from the Swedish Research Council.
We would like to thank E.~R.~Anderson, S.~K.~Bogner, R.~J.~Furnstahl and K.~Hebeler for useful discussions.

\end{acknowledgments}


\end{document}